\documentclass{article}

\usepackage{psfig}

\def\beq{\begin{equation}}
\def\eeq{\end{equation}}
\def\bce{\begin{center}}
\def\ece{\end{center}}
\def\bea{\begin{eqnarray}}
\def\eea{\end{eqnarray}}
\def\ben{\begin{enumerate}}
\def\een{\end{enumerate}}

\def\brr{\begin{array}}
\def\err{\end{array}}

\let\mbox=\hbox
\def\al{\alpha}
\def\be{\beta}
\def\ga{\gamma}

\def\na{\nabla}

\def\ph{\varphi}

\begin{document}

%\markboth{G. de Berredo-Peixoto}
%{On the 1-loop calculations of softly broken 
%fermion-torsion theory}

%\catchline{}{}{}{}{}
%
%%%%%%%%%%%%%%%%%%%%%%%%%%%%%%%%%%%%%%%%%%%%%%%%%%%%%%%%%%%%%%%%%%%%

\begin{center}
{\Large {\bf On the 1-loop calculations of softly broken fermion-torsion 
theory in curved space using the St\"uckelberg procedure}} 
\vspace{4mm}

{\large G. de Berredo-Peixoto}  
\vspace{3mm}

Departamento de F\'{\i}sica, ICE, Universidade 
Federal de Juiz de Fora, Campus universit\'ario\\
Juiz de Fora, MG 36036-330 Brazil \\
guilherme@fisica.ufjf.br
\end{center} 
%\vspace{2mm}

%\begin{history}
%\received{Day Month Year}
%\revised{Day Month Year}
%\end{history}

\begin{abstract}
The soft breaking of gauge or other symmetries is the typical
Quantum Field Theory phenomenon. In many cases one can apply 
the St$\ddot{\rm u}$ckelberg procedure, which means introducing 
some additional field (or fields) and restore the gauge symmetry. 
The original softly broken theory corresponds to a particular 
choice of the gauge fixing condition. In this paper we use
this scheme for performing quantum calculations for 
fermion-torsion theory, softly broken by the torsion mass in 
arbitrary curved spacetime. \\
\end{abstract}


 %%% Quantum field theory in curved spacetime
 %%% Renormalization
 %%% Gauge field theories
 %%% Symmetry and conservation laws

\section{Introduction}	

Softly broken gauge symmetries are frequently an important property 
in Quantum Field Theory (QFT). One example is supersymmetry, 
which must be (most likely softly) broken in order to address the phenomenological applications and eventually experimental 
tests \cite{softSUSY}. 
Another interesting application of the softly symmetry breaking is 
the effective QFT approach to the propagating 
torsion \cite{betor,guhesh,torsi}. 
The completely antisymmetric 
component of torsion can be described by the dual axial vector 
coupled to fermions through the axial vector current. The presence 
of the symmetry breaking mass of the axial vector is required for 
the consistency of the effective theory in the low-energy sector.
Indeed, the massive couterterm shows up at 1-loop level. 

In many cases, one is interested not only in the 
classical aspects of the theory, but also in the derivation 
of quantum corrections. The subject of the present paper is the calculation of
1-loop effective action for the softly broken gauge theory of
propagating torsion in curved 
space-time. Here, the kinetic term and the interactions terms in 
the classical action are gauge invariant while the massive terms 
are not. Consequently, the standard methods for evaluating the
effective action face serious technical difficulties. 
As a strategy, we shall apply the St$\ddot{\rm u}$ckelberg procedure \cite{stuck},
that is, we are going to restore the gauge symmetry by introducing 
an extra field or a set of fields. More details and applications of
the method to models with softly broken gauge symmetry in
curved spacetime can be found in Ref. \cite{bugui}.

We are going to show that
our approach means much 
simpler and more efficient calculation of quantum corrections. The 
difference is especially explicit for the massive torsion-fermion 
system which was originally elaborated in Ref. \cite{guhesh}. The present 
method provides an independent verification of our previous result 
in Ref. \cite{guhesh} and also enables one to perform the calculations in 
an arbitrary curved space-time, something that was impossible in the 
framework used in Ref. \cite{guhesh}. 

\section{Massive softly broken torsion field coupled 
to fermion}

Torsion $\,T^\alpha_{\;\beta\gamma}\,$ is an independent (along 
with the metric) quantity describing the spacetime manifold. 
It is defined by the relation (see, e.g., Refs. \cite{torsi,hehl} 
for introduction)
$$
{\Gamma}^\alpha_{\;\beta\gamma} -
{\Gamma}^\alpha_{\;\gamma\beta} =
T^\alpha_{\;\beta\gamma}\,.
$$
It proves useful to divide torsion into three irreducible 
components $\,T_{\mu},\,S_{\mu},\,q_{\alpha\beta\mu}\,$  
as already known in literature. The interaction with the Dirac 
fermion is described in a quantum consistent way by the action
for the theory of effective fermion-torsion system \cite{betor} 
(see also Refs. \cite{guhesh,torsi}),
\beq
S_{tf} = \int d^4x \sqrt{g}\,\left\{ - \frac14\,S_{\mu\nu}^2
- \frac12\,M^2 S^2_\mu \,\,+\,\, i\bar{\psi}\gamma^\mu 
\big( \na_\mu + i \eta\gamma^5S_\mu\big)\psi 
+ m\bar{\psi}\psi \,\right\}\,.
\label{t5}
\eeq
Here $S_{\mu\nu}=\partial_\mu S_\nu - 
\partial_\nu S_\mu$, $M$ is 
the torsion mass, we consider only one non-vanishing 
component of torsion, $T_{\al\be\ga}= - \frac{1}{6} \varepsilon_{\alpha\beta\ga\mu}
\,S^{\mu}$, and $\na_\mu$ is the covariant derivative without
torsion.

To calculate the 1-loop effective action for this model, 
one has to apply the generalized method of 
Schwinger-DeWitt \cite{bavi85} in the transverse vector space, as
was done in Ref. \cite{guhesh}.
Following the approach discussed in Ref. \cite{bugui},
one can apply the St$\ddot{\rm u}$ckelberg procedure by
introducing a new scalar field, $\ph$, and restoring the gauge 
symmetry in the following way:
\bea
S_{tf}^\prime 
&=& \int d^4x \sqrt{g}\, \Big\{ - \frac14\,S_{\mu\nu}^2
+ \frac12\,M^2\Big(S_\mu - \frac{\partial_\mu \ph}{M}\,\Big)^2
\nonumber
\\
&+&  i\bar{\psi}\gamma^\mu 
\big(\na_\mu + i \eta_1\gamma^5S_\mu\big)\psi 
\,\,+\,\, m \,\bar{\psi}\,
\exp \left(\frac{2i\eta\,\ga^5\ph}{M}\right)\,\psi 
\Big\}\,,
\label{t555}
\eea
The gauge symmetry must be supplemented 
by $\ph \to \ph^\prime = \ph - M\be$. The original theory 
(\ref{t5}) is restored when we use the gauge fixing condition 
$\ph=0$.

\section{One-loop effective action and quantum \\ 
(in)consistency}

In order to obtain the one-loop divergences for the original 
theory (\ref{t5}), one has to put $\varphi = 0$ in the 
general expression for the divergences of theory (\ref{t555}),
which can be computed by the standard Schwinger-DeWitt method.
Then the final result reduces to 
\bea
\Gamma^{(1)}_{{\rm div}} & = &
-\frac{\mu^{n-4}}{(4\pi)^2(n-4)} \int d^nx\sqrt{g}\left\{
4\eta^2 m^2 S^\mu S_\mu - \frac13 \eta^2 S_{\mu\nu}^2 
+ 4i\eta^2\frac{m^2}{M^2}\bar{\psi}\ga^\mu {\cal D}_\mu^{\ast}\psi
+ \right. \nonumber \\
& + & \left.
2i\eta^2\,{\bar \psi}\ga^\mu {\cal D}_\mu\psi
+ \left(\frac{8\eta^2 m^3}{M^2} - 4\eta^2 m \right)\bar{\psi}\psi
+ \frac{2\eta^2 m}{3M^2}\bar{\psi}\,R\,\psi 
+ \frac{8\eta^4 m^2}{M^4}(\bar{\psi}\psi)^2  
\right\}\,, \nonumber \\
\label{r1}
\eea 
where ${\cal D}_\rho = \na_\rho + i\eta\ga^5 S_\rho$ and 
${\cal D}^{\ast}_\rho = \na_\rho - i\eta\ga^5 S_\rho$. 

It is worth mentioning that the above result is more
general than the result of Ref. \cite{guhesh}. Indeed,
it is valid in curved spacetime, where a new 
non-minimal coupling with curvature shows up. Of course this
term is relevant for dynamics of Dirac particles in the curved
background, but the theory contains the 
$(\bar{\psi}\psi)^2$-term which has non-trivial consequences.
In fact, at 2-loop level, this term is responsible for
appearance of the Feynman diagrams drawn in Fig. 1.

\begin{figure}[ph]
\centerline{\psfig{file=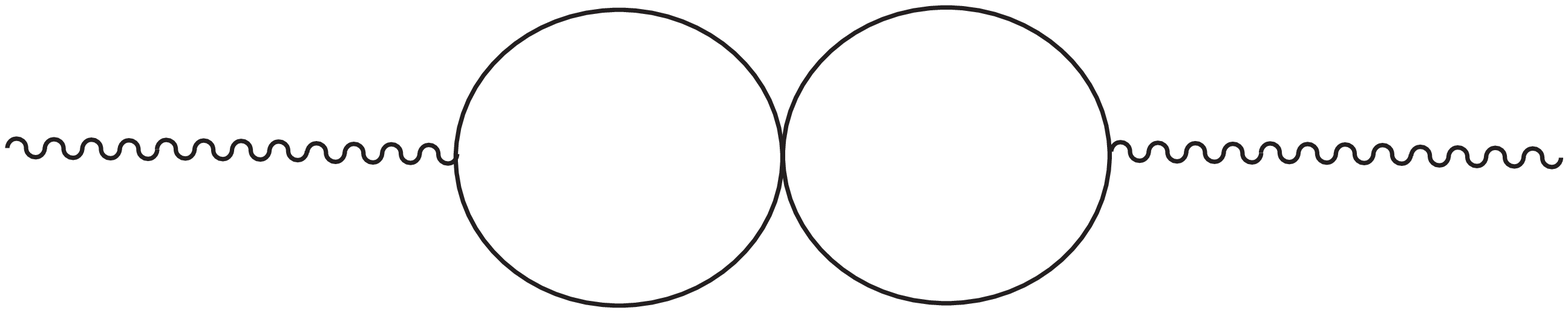,width=4.8cm}$\;\;$
\psfig{file=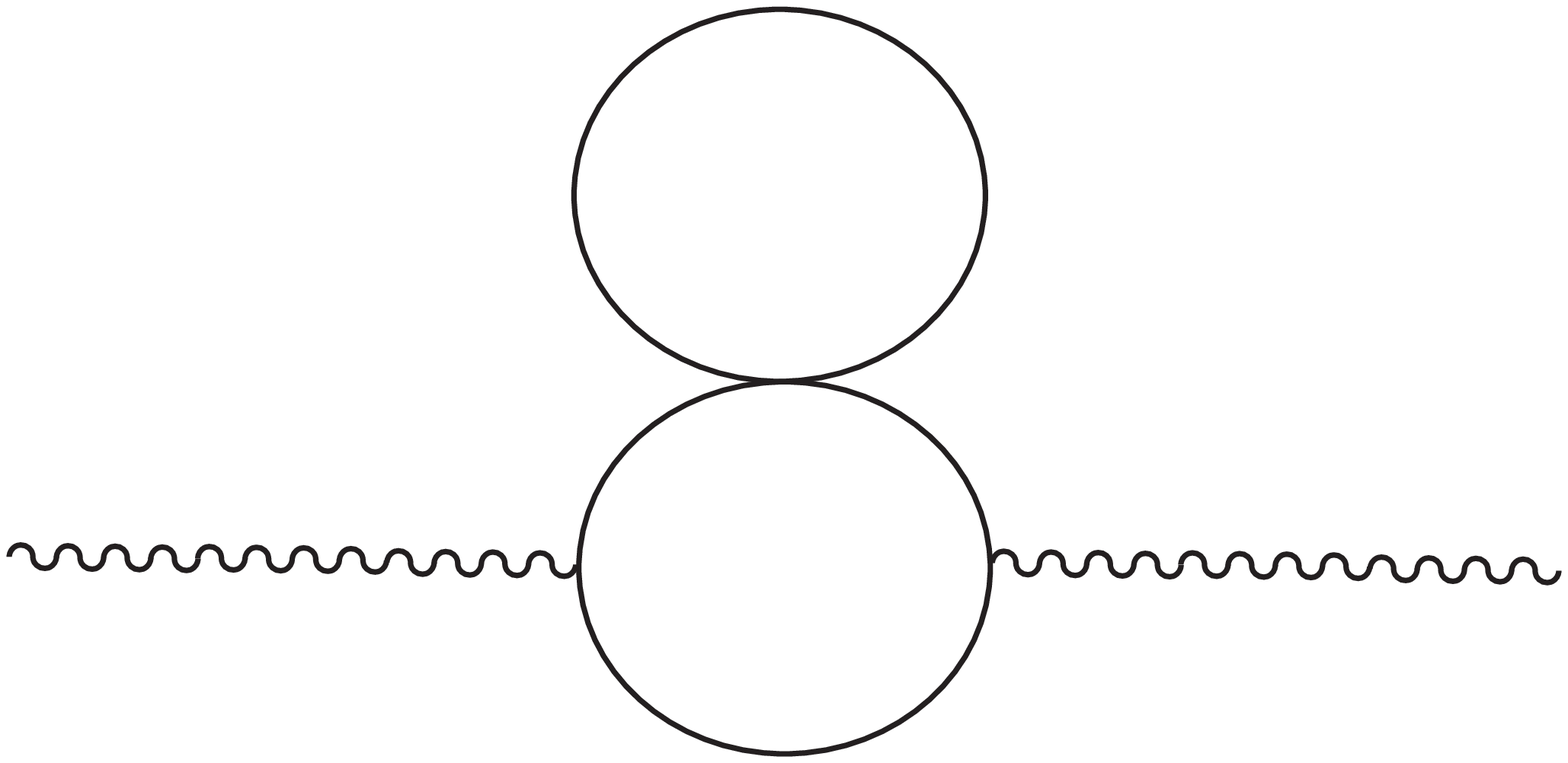,width=4cm}}
\vspace*{5pt}
\caption{2-loop diagrams with quartic fermionic vertices. The
wavy lines correspond to torsion propagator, and the others to
the fermion one. \label{f1}}
\end{figure}

Detailed calculation of these diagrams reveals the appearance
of the $(\partial_\mu S^\mu)^2$-type counterterm, which introduces
longitudinal degrees of freedom breaking unitarity. This undesireble
contribution can not be compensated by another 2-loop diagrams,
unless some artificial fine-tunning between different coupling
constants takes place.

Even if theory (\ref{t5}) is not consistent at the quantum level,
it can pehaps mimic some fundamental theory, as an effective 
theory. In this sense, it would be interesting to study the
phenomenological consequences of the coupling term, 
$g R\bar{\psi}\psi$. For instance, this term seems to introduce
some kind of modified fermion mass, giving rise to an interesting
non-trivial effect on the mass renormalization.

\subsection*{Acknowledgments}
The author is grateful to Prof. I.L. Buchbinder
and Prof. I.L. Shapiro for fruitful discussions, and
also acknowledges support from CNPq, 
FAPEMIG and FAPES.

\end{document}